\documentclass[a4paper]{article}

\usepackage{INTERSPEECH2020}
\usepackage{xcolor}
\usepackage{url}
\usepackage{multirow}

\title{Voice Conversion by Cascading  Automatic Speech Recognition and Text-to-Speech Synthesis with Prosody Transfer}
\name{Jing-Xuan Zhang$^1$, Li-Juan Liu$^2$, Yan-Nian Chen$^{12}$, Ya-Jun Hu$^2$, Yuan Jiang$^2$, Zhen-Hua Ling$^1$, Li-Rong Dai$^1$\thanks{This work was partially funded by the National Nature Science Foundation of China under Grant 61871358.}}
\address{
  $^1$National Engineering Laboratory of Speech and Language Information Processing,\\
  University of Science and Technology of China, P.R.China\\
  $^2$iFLYTEK Research, iFLYTEK Co., Ltd.}
\email{nosisi@mail.ustc.edu.cn, \{ljliu,yanchen5,yjhu,yuanjiang\}@iflytek.com, \{zhling,lrdai\}@ustc.edu.cn}

\begin{document}

\maketitle
\begin{abstract}
With the development of automatic speech recognition (ASR) and
text-to-speech synthesis (TTS) technique, it's intuitive to construct a
voice conversion system by cascading an ASR and TTS system.
In this paper, we present a ASR-TTS method for voice conversion, which used
iFLYTEK ASR engine to transcribe the source speech into text and
a Transformer TTS model with WaveNet vocoder to synthesize the converted
speech from the decoded text.
For the TTS model, we proposed to
use a prosody code to describe
the prosody information other than
text and speaker information contained in speech.
A prosody encoder is used to extract the prosody code.
During conversion, the source prosody is transferred to converted speech by conditioning the
Transformer TTS model with its code.
Experiments were conducted to demonstrate the effectiveness of our proposed method.
Our system also obtained the best naturalness and similarity in the mono-lingual task of Voice Conversion Challenge 2020.

\end{abstract}


\begin{figure*}[t]
\centering
\includegraphics[width=0.8\textwidth]{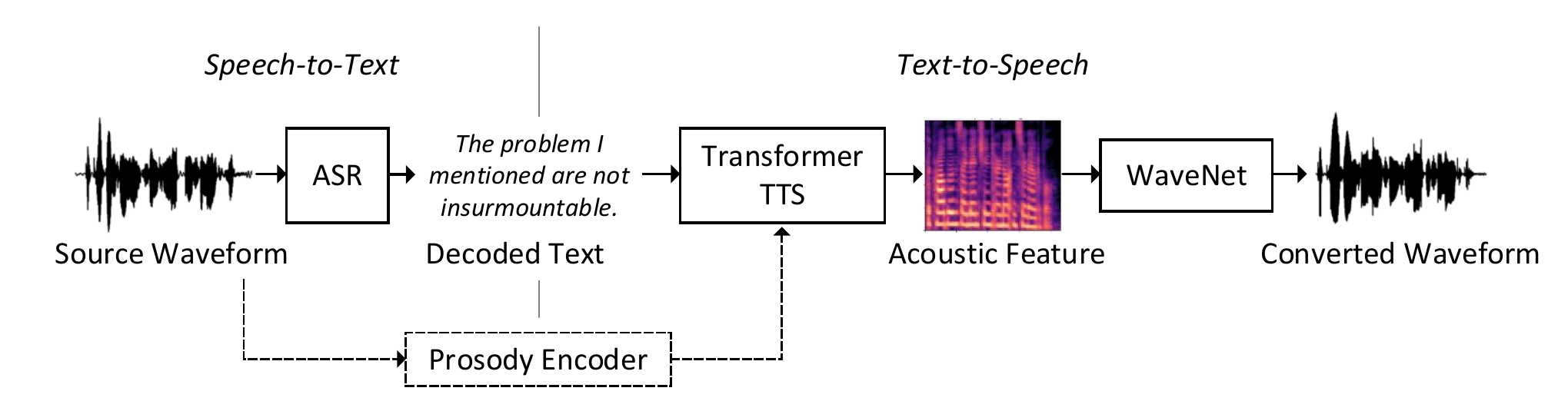} 
\caption{Overview of our proposed method for voice conversion. The prosody encoder part is optional.}
\label{fig:fig2}
\end{figure*}

\section{Introduction}
\label{sec:intro}
The objective of voice conversion (VC) is to modify
certain aspects of the speech, such as accent or speaker identity, while preserving the linguistic content unchanged during this process \cite{Childers1989Voice,Childers1985Voice}. The conversion of speaker identity might be the most popular task of voice conversion, in which the utterance of a source speaker is processed to make it sounds like a target speaker. This work will also focus on the speaker conversion. Voice conversion has many practice applications, such
as entertainment, personalized text-to-speech synthesize and voice anonymization.

Voice conversion has been well studies in recent decades.
A conventional voice conversion system is based on the parallel training data,  i.e., the same utterances are spoken by both the source and the target speaker.
Joint-density Gaussian
mixture model (GMM) \cite{Kain1998Spectral,Toda2007Voice}, deep neural network (DNN) \cite{Desai2009voice,Desai2010Spectral} and recurrent neural nework (RNN) \cite{Sun2015Voice,nakashika2015voice} and sequence-to-sequence (seq2seq) model \cite{8607053,tanaka2019} have been used for VC.
Compared to the parallel VC,
non-parallel VC is not constrained by the
parallel training data.
Many non-parallel VC methods have been proposed, such as varitional auto-encoder (VAE) based VC \cite{hsu2016voice,hsu2017voice}, CycleGAN \cite{kaneko2017parallel} and StarGAN \cite{kameoka2018stargan} based VC.
Among those methods, recognition-synthesis based VC is one of the most popular \cite{sun2016phonetic,miyoshi2017voice,ljliu2018wav,liu2018voice}. In this method, an automatic speech recognition (ASR) model is first trained
to learn a mapping function from speech to text. Then, the
linguistic descriptions are extracted by the ASR model. They are typically outputs from
a certain layer of the ASR model, such as posteriorgrams (PPGs) \cite{sun2016phonetic} or content-related features \cite{ljliu2018wav}.
Then, a synthesis model is trained to predict the target acoustic features
using those linguistic descriptions as inputs.
During the conversion stage,
the source acoustic features are first encoded into the linguistic descriptions,
which are then decoded to the target acoustic features.
Our previous method using content-related features obtained the best performance in Voice Conversion Challenge 2018 \cite{ljliu2018wav}.
Although its success, in those method, the source speaker information may not be completely separated from the PPGs or content-related features, which may harm the similarity of converted voice. Also,
the model is constrained to perform the conversion frame by frame,
since the PPGs or content-related features are frame level features. Therefore it's difficult to apply the
seq2seq modeling, which has been proven the advantage in a variety of speech generation tasks \cite{8607053,liu2020voice,wang2017tacotron,li2019neural}.

Recent years, automatic speech recognition and text-to-speech synthesize have become
mature technique with research and development efforts from speech process communities.
Under some circumstance, the automatic speech recognition has achieved accuracy levels comparable
to human transcribers \cite{stolcke2017}.
By using seq2seq acoustic modeling with neural vocoder \cite{denoord2016wavenet}, the synthesized speech is close to human quality \cite{shen2017natural,li2019neural}.
Given those facts,
it's intuitive to construct a voice conversion system by cascading a
ASR and TTS system directly, 
which is also a kind of 
recognition-synthesis based VC method. In this method,  rather than PPGs or content-related features, text is used as the intermediate description to bridge different speakers, which only contains the sematic information.
The iFLYTEK ASR engine and a Transformer TTS model were used in our experiments.
In order to model the prosody information
in addition to text and speaker identity,
we further extracted a global prosody code 
as input to the TTS model.
During conversion, the prosody code from the
source utterance is transferred and used to condition the
target Transformer TTS model \cite{li2019neural} for generating more natural utterance. 
The overall scheme of our method is depicted in Figure~\ref{fig:fig2}.


In our experiments section, we compared our proposed method with
the content-related features based baseline.
Our method achieved the best performance on the mono-lingual task of Voice Conversion Challenge 2020, in terms of both naturalness and similarity. It's also noticeable our system has no statistical significant
difference with the natural speech in terms of
similarity.




\section{Background}
\label{sec:back}

\begin{figure}[t]
\centering
\includegraphics[width=0.9\columnwidth]{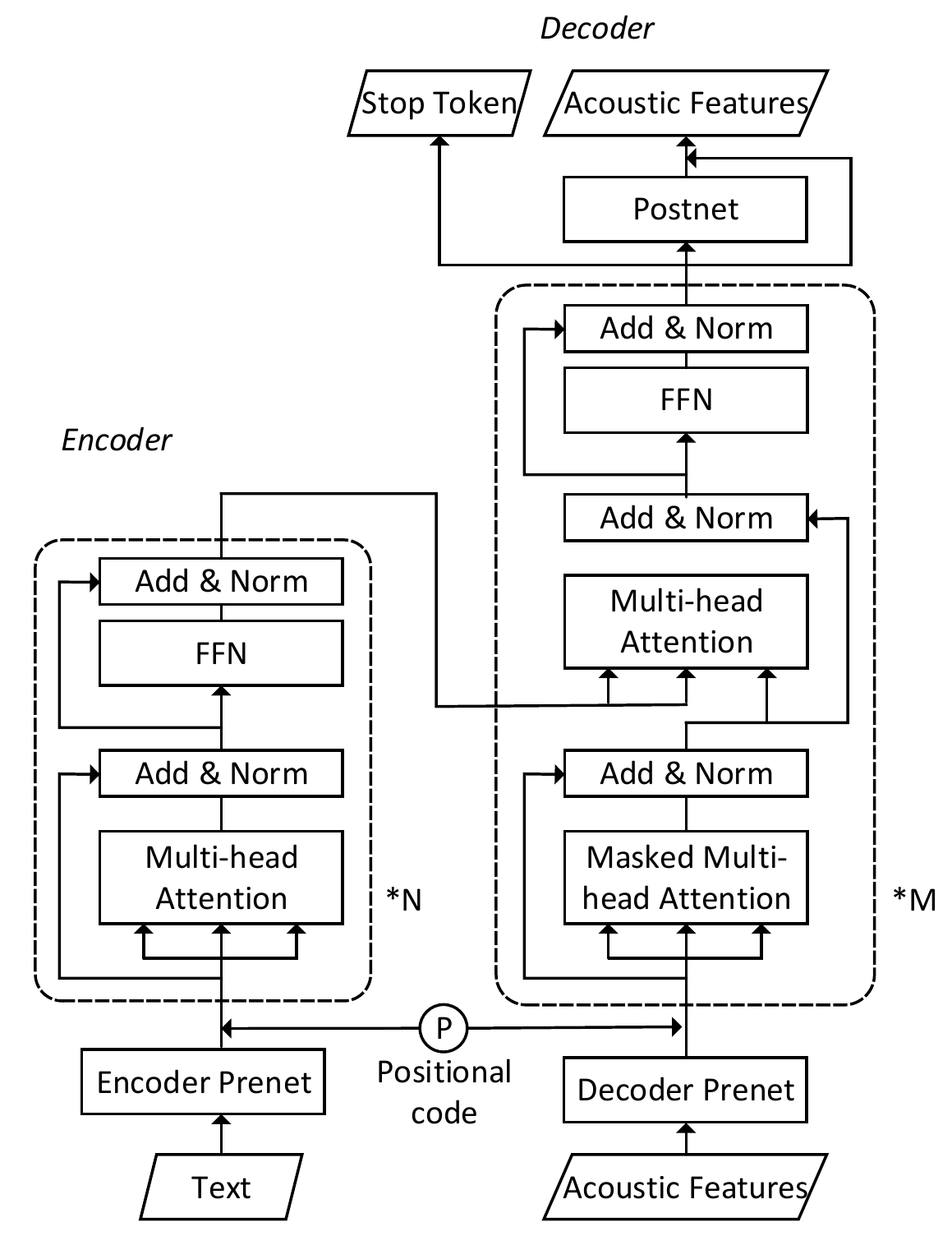} 
\caption{Structure overview of Transformer TTS.}
\label{fig:fig1}
\end{figure}

Transformer TTS is a seq2seq model that has been proposed for text-to-speech synthesis \cite{li2019neural}.
Transformer TTS mainly rely on the multi-head attention
mechanism \cite{transformer17} instead of convolutions in CNN or recurrences in RNN
for modeling sequential data.
It is composed of a text encoder and a decoder module as illustrated in Figure~\ref{fig:fig1}. 
The text is first processed with a encoder Prenet then added with triangle positional embeddings. Then they are passed through
stacks of encoder blocks. Each block contains two sub-layers,
including a multi-head self-attention mechanism and
a position-wise fully connected network.
Also, residual connection followed by layer normalization is employed after each sub-layer.
The decoder predicts the next frame auto-regressively, consuming the previously generated acoustic frame
and the encoder outputs as the inputs.
The acoustic frame is first passed through a
decoder Prenet then fused with the positional code by adding.
Then it's passed through a stacks of decoder blocks.
Compared to the encoder block,  the decoder block inserts
an extra sub-layer that attends to the encoder outputs.
Also, the self-attention in the decoder is masked to ensure the
causality, i.e. the decoder should only make use of the history
information each step.
At last, in order to refine the acosutic features' reconstruction, a CNN based Postnet is employed to
produce a residual, which is added with initial outputs.
Also, a stop token is predicted to determine the end of
generation process. For the detail of Transformer TTS, it suggested that the reader refer to the original paper \cite{li2019neural}.

Compared to the RNN based seq2seq TTS model, such as Tacotron, there're two advantages of Transformer TTS.
First, without using any
recurrent units, the computation of
Transformer TTS is highly parallel during the training time, leading to higher training efficiency. Second, by using multi-head attention mechanism,
two frames with any distance in a sequence can be
associated in one step, thus
the long-range dependency can be easily captured.



\section{Proposed Method}
\label{sec:prop}
\begin{figure}[]
\centering
\includegraphics[width=\columnwidth]{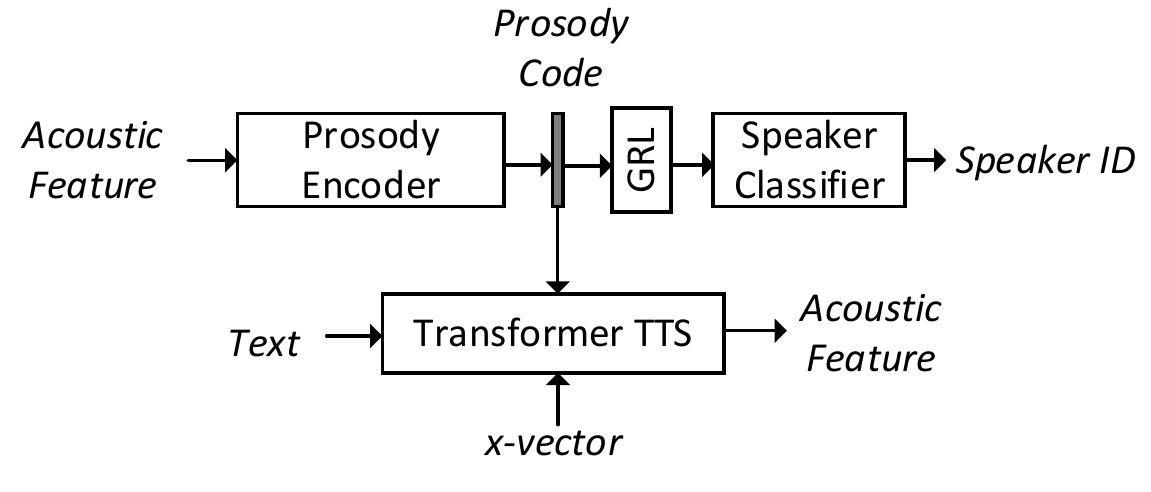} 
\caption{Multi-speaker Transformer TTS with prosody code conditioning. GRL denotes the gradient reversal layer.}
\label{fig:fig3}
\end{figure}


\subsection{ASR model}
The ASR model is used to decode the text from source utterances, which determines the content of the converted
utterances. By definition of voice conversion task,
linguistic content should not be changed during this process.
Therefore, it's crucial that the ASR model to recognize
the text information as much accuracy as possible.
For the ASR model, we compared two methods, including the commercial iFLYTEK ASR engine and an  open-source ASR model based on ESPnet \cite{espnet}. And results were reported in
our experiments section.

\subsection{TTS model}
We extended  Transformer TTS to the multi-speaker
condition by using speaker x-vectors \cite{snyder2018x} as additional inputs to its decoder.
The x-vectors are projected by linear transformation then added to each frame of encoder outputs.
Another difference with the original Transformer TTS paper is that we don't use the encoder Prenet, since
no explicit improvement was found in our preliminary experiments.
 The amount of target training data is usually small in voice conversion task. Thus
the model often suffers from over-fitting effect if
no external dataset is exploited. Therefore,
in order to improving the generalization capacity of the model,
it is first pre-trained on a multi-speaker dataset, and then adapted to the target speaker by finetuning.

\subsection{Prosody transfer}
We assume that the speech signal contains
not only the text and speaker identity information, but also the prosody information. Therefore,
a prosody code is further provided to the synthesis model to describe the prosodic ingredients of the speech.
In our method,
a prosody encoder is employed to extract
a global prosody code for each utterance
as shown in Figure~\ref{fig:fig3}.
The Transformer TTS model is further conditioned on the prosody code. Specifically,
the code is projected by linear transformation then added to each frame of encoder outputs.
During conversion, the source prosody code is transferred and
used to synthesize the converted utterance to make it more natural sounded.
However,
because that the TTS model is pre-trained on
the multi-speaker dataset, the desired prosody information
can be easily entangled with the speaker identity information. Therefore,
we further adopted a speaker adversarial loss
to regularize the extracted code to be speaker-independent.
To this end, a speaker classifier is adopted
to predict the speaker identity from the
prosody code. While the prosody encoder is optimized with the opposite objective to
lower the classifier's accuracy of prediction.


In our implementation,  a gradient reversal layer is inserted between
the prosody code and the classifier.
The speaker classifier is updated four times for each update of the prosody encoder. It is critical
to keep the classifier close to the optimal so that it can backpropagate useful gradients rather than noises.
Our prosody encoder is only updated during the multi-speaker pre-training stage while it's frozen during fine-tuning on the target speaker.


\section{Experiments}
\label{sec:exp}

\begin{table}[t]
    \caption{Character Error Rate (CER) and Word Error Rate (WER) of iFLYTEK ASR engine and ESPnet based ASR model.}
    \label{tab:tab1}
    \centering
    \begin{tabular}{c | c | c }
        \hline
        \hline
             & CER (\%) & WER (\%) \\
             \hline
        iFLYTEK & 0.958  & 2.954  \\
         ESPnet & 2.415  & 7.0l57 \\

        \hline
         \hline
    \end{tabular}
\end{table}

\subsection{Experiment conditions}
Dataset from the mono-lingual task of Voice Conversion Challenge 2020
 was adopted, which contained 4 source English speakers (SEF1, SEF2, SEM1 and SEM2) and 4 target English speakers (TEF1, TEF2, TEM1 and  TEM2). Each speaker contained 70 utterances for training, 20 of which were parallel
for source and target speakers and the remaining were non-parallel.
Our model only relies on the target utterances for model adaptation. Therefore, the non-parallel part of source utterances were used as our internal evaluation set.
For evaluation in Voice Conversion Challenge 2020, 25 test utterances were further provided for each source speaker. And they were converted to each target speaker.
LibriTTS corpus\footnote{\url{http://www.openslr.org/60/}} was adopted for pretraining our model. We used about 460 hours training data from 1150 speakers.

For ASR model, we compared the iFLYTEK ASR engine and
a open-source ASR model\footnote{\url{https://drive.google.com/file/d/1BtQvAnsFvVi-dp_qsaFP7n4A_5cwnlR6/view?usp=drive_open}}
based on ESPnet \cite{espnet} and the results on our internal evaluation set were presented in Table~\ref{tab:tab1}. We observed that the iFLYTEK ASR engine achieved better performance than the ESPnet based one. Therefore, the former was adopted in our experiments.

The Transformer TTS model contained 6 blocks for both encoder and decoder. The number of attention head was 4 and the layer width was 1536. The architecture of the prosody encoder followed the reference encoder of GST-Tacotron \cite{wang2018style}, producing a 128-dimensional prosody code for each utterance. The speaker classifier was a $3 \times 512$ DNN. 512-dimensional x-vectors were extracted by Kaldi toolkit\footnote{\url{https://kaldi-asr.org/}}.  For acoustic features, 80-dimensional Mel-spectrograms were extracted with Mel filters spanning from 80 Hz to 7600 Hz. We used Adam optimizer and Noam scheduler \cite{transformer17} for adjusting the learning rate. The batch size was 120, which was later set as 16 during finetuning. At last, WaveNet vocoder was trained to model 24kHz-16bit waveforms for each target speaker.

\subsection{Comparison with the baseline}
We first compared our proposed ASR-TTS method without using
prosody code to a baseline method, which was an improved version
of N10 system in Voice Conversion Challenge 2018, denoted by \textbf{VCC2018+}. Compared to N10 system, this baseline adopted a new auto-regressive synthesis model.
Also, a higher quality WaveNet vocoder was adopted. For the details of this model, the reader can refer to another our paper for the cross-lingual task of Voice Conversion Challenge 2020.

\begin{table}[t]
    \caption{Mean opinion score (MOS) and 95\% confidence interval of our proposed method and the baseline. Natural denotes the natural target speech.}
    \label{tab:tab2}
    \centering
    \begin{tabular}{c | c | c }
        \hline
        \hline
             &  Naturalness &Similarity \\
             \hline
         VCC2018+ & 3.781$\pm$0.070 & 3.610$\pm$0.076 \\
         proposed & 3.802$\pm$0.067 & 3.739$\pm$0.074 \\
         Natural & 3.815$\pm$0.077 & 3.703$\pm$0.08 \\
        \hline
         \hline
    \end{tabular}
\end{table}

We conducted the subjective listening test in terms of
both naturalness and similarity. 40 utterances for each method were randomly selected from our internal evaluation set.
We used Amazon Mechanical Turk\footnote{\url{https://www.mturk.com/}} platform and at least 30 native listeners participated in each of our experiments.
They were required to used headphones during the test and the stimuli were presented in random order. Table~\ref{tab:tab2} summarized the results. We observed from the table that our proposed method achieved higher naturalness and similarity than the VCC2018+ baseline. And it even obtained similar results to the natural target speech.


\subsection{Using prosody transfer strategy}

\begin{figure}
    \centering
    \includegraphics[width=0.75\columnwidth]{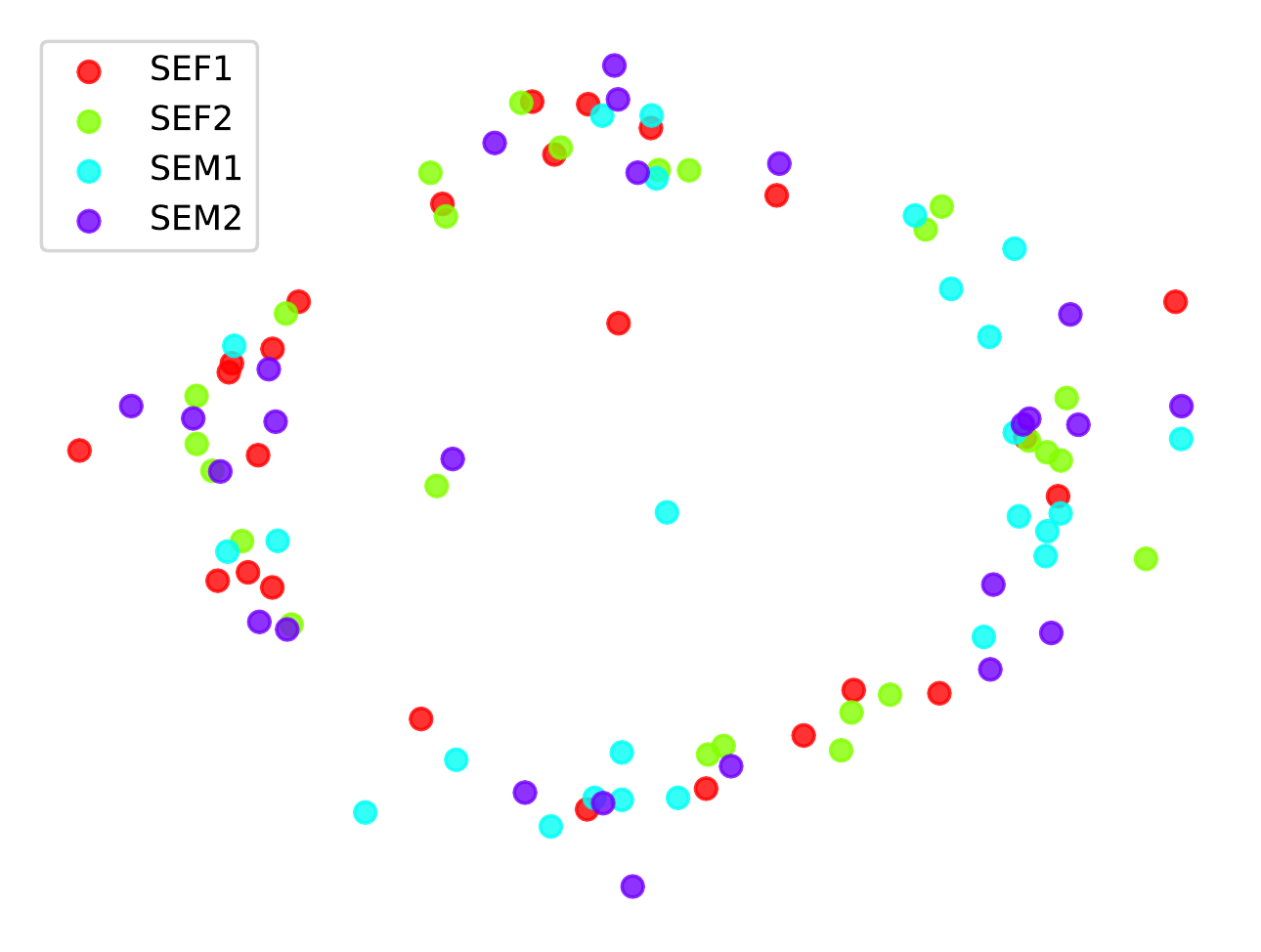}
    \caption{Visualization of prosody code from four source speakers. Each point denotes a prosody code from an utterance.}
    \label{fig:fig4}
\end{figure}

In order to analyze the latent space of prosody code, they were
extracted from four source speakers, i.e., SEF1, SEF2, SEM1 and SEM2, then  projected to 2 dimension by t-SNE for visualization. The 25 test utterances for each speaker were used and the results were presented in Figure~\ref{fig:fig4}.
We observed from this figure the prosody code from each speaker equally distributed in the same latent space,
which demonstrated the speaker-indepentent
property of the prosody code.

Based on the preliminary experiments, we heuristically scaled the prosody code with a factor of 1.5 during conversion. And the proposed method with and without prosody code were compared by subjective listening test. 10 sentences from internal evaluation set were randomly  selected for each target speaker, resulting
a total of 40 sentences for evaluation. 
The results were grouped by the target speaker as presented in Table~\ref{tab:tab3}. As we can see
from the table, the proposed method with and without prosody code obtained close naturalness and
similarity score. Averagely, \emph{w pc} method obtained slightly better naturalness but
slightly lower similarity than \emph{w/o pc} method. For the target speaker TEF2, \emph{w pc} method achieved slightly better results in both naturalness and similarity. It should be notice that our \emph{w/o pc} method
already achieved comparable results with the natural target speech, as shown in previous section. Therefore,
there's no clear advantage of using prosody transfer strategy in our experiments.
To explore this strategy with dataset containing expressive utterances will be an interesting research direction in the future.

\begin{table}[t]
    \caption{Mean opinion score (MOS) and 95\% confidence interval of the proposed method with (w pc) and without (w/o pc) using prosody code.}
    \label{tab:tab3}
    \centering
    \begin{tabular}{c | c | c | c }
        \hline
        \hline
        Target speaker & Method  & Naturalness & Similarity \\
             \hline
         \multirow{2}{*}{TEF1} & \emph{w/o pc} & 3.673$\pm$0.188 & 3.593$\pm$0.193 \\
              & \emph{w pc} & 3.713$\pm$0.183 & 3.553$\pm$0.189 \\
         \multirow{2}{*}{TEF2} & \emph{w/o pc} & 3.705$\pm$0.163 & 3.771$\pm$0.165 \\
              & \emph{w pc} & 3.757$\pm$0.160 & 3.871$\pm$0.162 \\
         \multirow{2}{*}{TEM1} & \emph{w/o pc} & 3.729$\pm$0.162 & 3.843$\pm$0.163 \\
              & \emph{w pc} & 3.781$\pm$0.160 & 3.724$\pm$0.163 \\
         \multirow{2}{*}{TEM2} & \emph{w/o pc} & 3.711$\pm$0.162 & 3.632$\pm$0.178 \\
              & \emph{w pc} & 3.668$\pm$0.166 & 3.616$\pm$0.181 \\
         \hline
         \multirow{2}{*}{Average} & \emph{w/o pc} & 3.707$\pm$0.083 & 3.721$\pm$0.087 \\
              & \emph{w pc} & 3.733$\pm$0.082 & 3.704$\pm$0.086 \\
        \hline
         \hline
    \end{tabular}
\end{table}

\subsection{Voice Conversion Chanllenge 2020}

\begin{figure}[t]
    \centering
    \includegraphics[width=0.9\columnwidth]{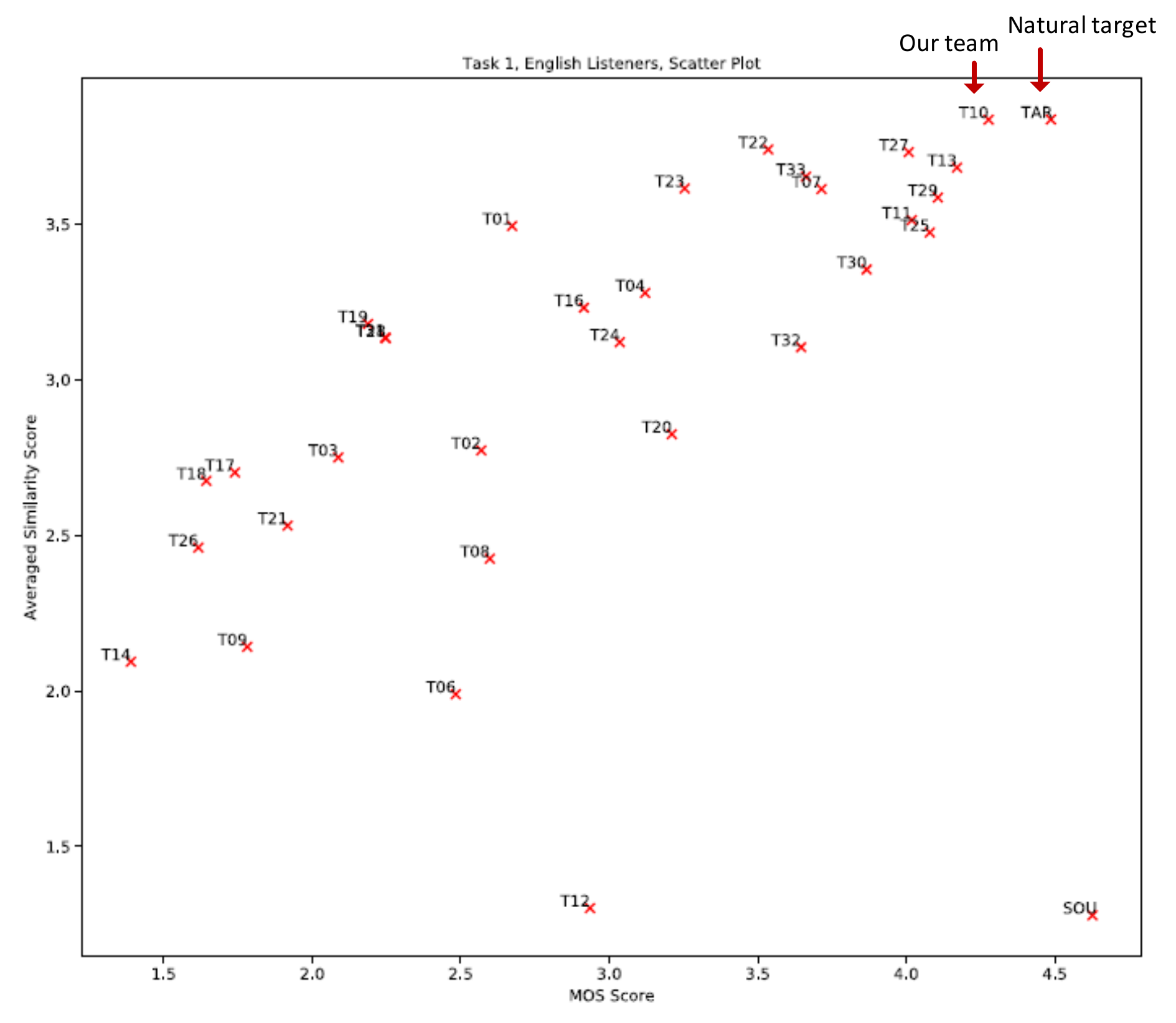}
    \caption{Average similarity score and naturalness MOS for different teams in Voice Conversion Challenge 2020.}
    \label{fig:fig7}
\end{figure}

Different conversion methods were selected according to their performance on the conversion pairs in Voice Conversion Challenge 2020\footnote{\url{http://www.vc-challenge.org/}}.
Based on the experimental results in previous section,
we adopted the proposed method with prosody transfer strategy when the target speaker was
TEF2. For SEM1-to-TEM1 and SEM1-to-TEM2, VCC2018+ method was used since the proposed ASR-TTS achieved similar results to it on those conversion pairs. For remaining conversion pairs, we used the proposed method without prosody transfer strategy.

The subjective listening results, which were collected from 68 native English speakers,
were shown in Figure~\ref{fig:fig7}. As we can observed from this figure, our method achieved
the best performance among the  participating teams. Also, it's noticeable our method
achieved similar similarity score with the natural target speech. According to further results provided by the organizer, converted utterances by our method had no statistical significant difference
with the natural target speech in terms of similarity.

\section{Conclusion}
\label{sec:con}
We present a voice conversion method by cascading
a ASR and TTS module in this paper. iFLYTEK ASR engine is used
for decoding text from the source utterance and a Transformer TTS model is
trained for synthesizing the target speech. 
The Transformer TTS is pretrained on a multi-speaker dataset then finetuned
 on the target speaker in order to boosting its generalization capacity. A prosody transfer technique is further proposed,
in which a prosody code is extracted by a prosody encoder from the source
then used to condition the target TTS model.
Our experimental results showed the effectiveness of proposed method for voice conversion.
However, the ASR model is still imperfect. 
The speech recognition errors will lead to 
the change of the linguistic content of  converted speech.   
Investigating on how to minimize
those errors by using a mix of recognized text and 
the content-related features will be investigated in our future work.

\bibliographystyle{IEEEtran}
\bibliography{mybib}

\iftrue
\clearpage
\section{Appendix}
\subsection{Analysis on the content-related features}

\begin{table}[t]
    \caption{Mean opinion score (MOS) and 95\% confidence interval of VCC2018+ method using matched and mismatched inputs.}
    \label{tab:match_unmatch}
    \centering
    \begin{tabular}{c | c | c }
        \hline
        \hline
             & Naturalness & Similarity \\
             \hline
         \emph{matched} & 3.872$\pm$0.099 & 3.842$\pm$0.111 \\
         \emph{mismatched} & 3.816$\pm$0.099 & 3.723$\pm$0.110 \\
        \hline
         \hline
    \end{tabular}
\end{table}

Unlike  text,  the  content-related  features  contained  the source speaker information, therefore may harm the similarity of converted speech.  In order to prove this, we compared between inputting the target content features (i.e., the content features and acoustic model are \emph{matched}) and the source content features (i.e., the content features and acoustic model are \emph{mismatched}).  And the voice conversion is  mismatched condition. From Table~\ref{tab:match_unmatch}, we observed that the mismatched condition underperformed the matched one, especially in term of similarity.

\subsection{Analysis on the effect of prosody code}

\begin{figure}[t]
    \centering
    \includegraphics[width=0.8\columnwidth]{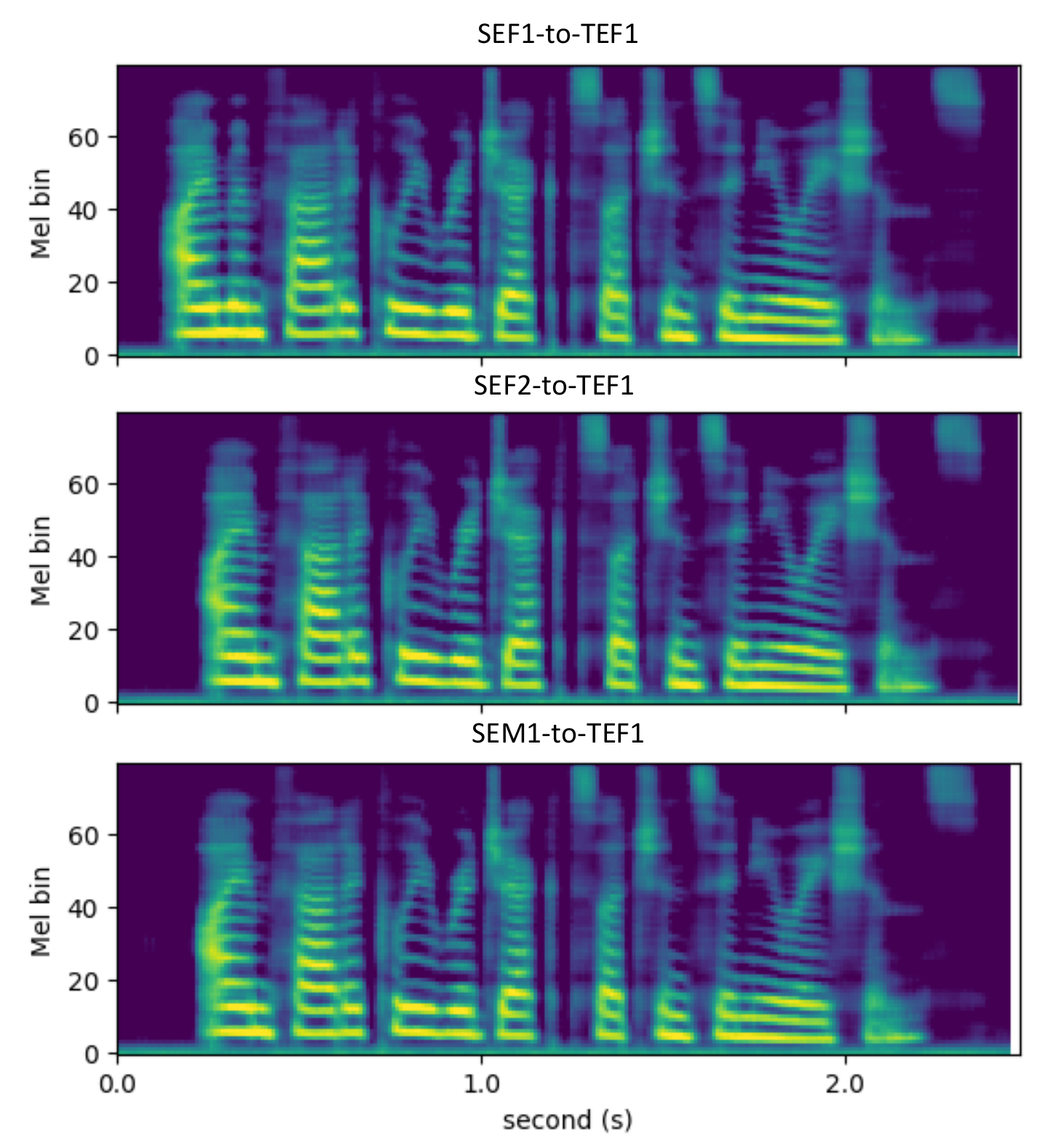}
    \caption{Mel-spectrograms of the converted target utterances with prosody code from different source speakers.}
    \label{fig:fig5}
\end{figure}

\begin{figure}[t]
    \centering
    \includegraphics[width=0.8\columnwidth]{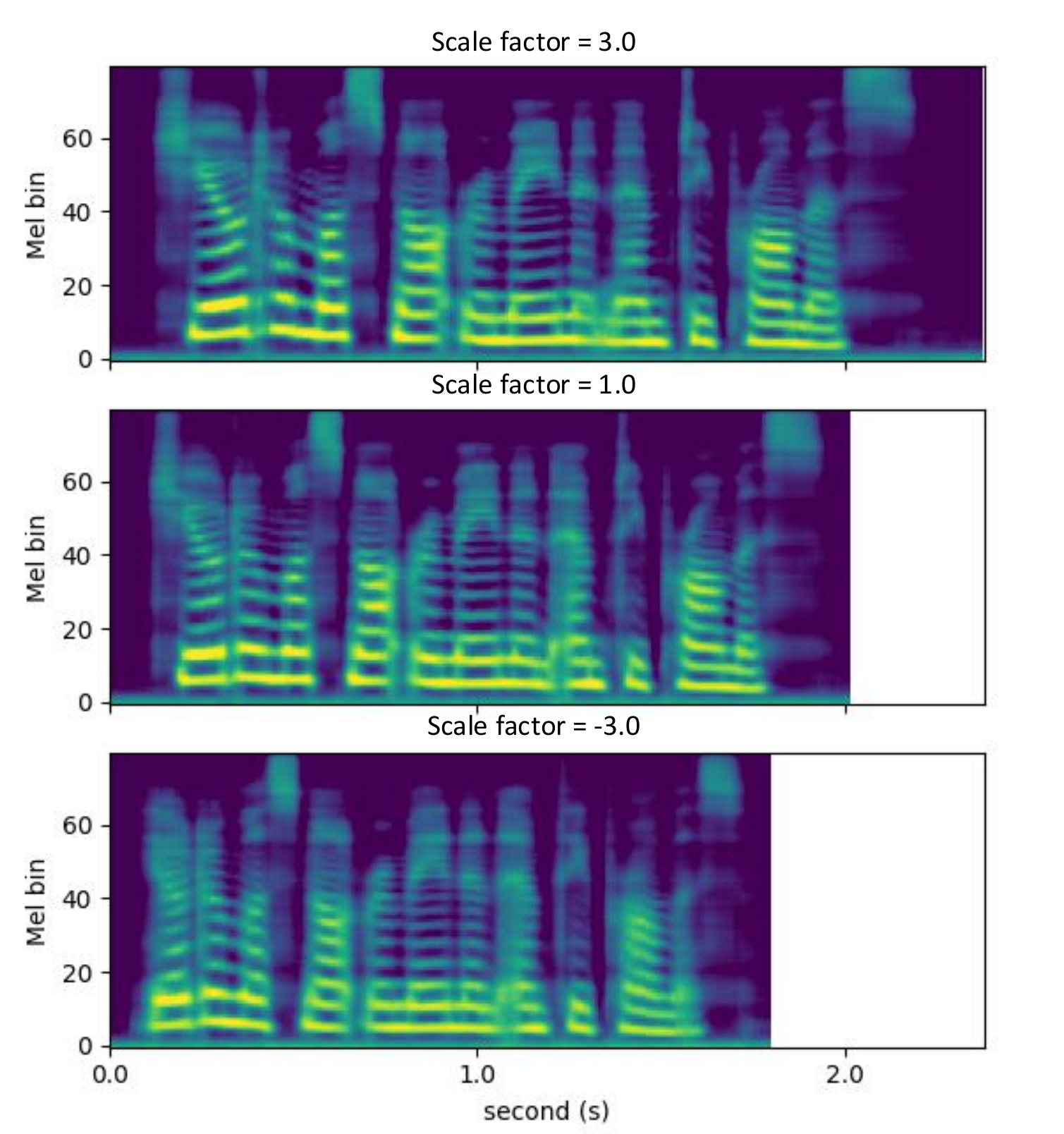}
    \caption{Mel-spectrograms of the converted target utterances with prosody code scaled by different factors.}
    \label{fig:fig6}
\end{figure}

The same target utterances were synthesized but with  the prosody code transferred from different source speakers.
Mel-spectrograms were shown in Figure~\ref{fig:fig5}. We observed from the figure that the synthesized Mel-spectrograms were similar to each other while slightly difference appeared, indicating the effectiveness of prosody code.

However, we found the control effect of prosody code was subtle thus hard to identify it.
In order to further investigate the
relationship of prosody code to the prosody aspect of generated speech,
we further conducted the experiments that first scale the prosody code by multiply it with a scale factor, then use it to control the synthesis process.
Mel-spectrogram of an utterance converted from SEF1 to TEF1 was presented in Figure~\ref{fig:fig6}.
From this figure, we observed that as we
scaled the prosody code from 3 to -3, the generated
Mel-spectrogram got more blurred, the speaker rate got faster and the pitch contour got more flatten.

\fi

\end{document}